\documentclass[12pt]{article}
\usepackage{latexsym}

\renewcommand{\theequation}{\arabic{section}.\arabic{equation}}

\begin{document}
\begin{titlepage}

\vspace{.7in}

\begin{center}
\Large
{\bf Embeddings in Non--Vacuum Spacetimes}
\\
\vspace{.7in}
\normalsize
\large{Edward Anderson$^1$  and James E. Lidsey$^2$}
\\
\normalsize
\vspace{.4in}
{\em Astronomy Unit, School of Mathematical Sciences, \\
Queen Mary, University of London, Mile End Road, London E1 4NS, U.K. }
\vspace{.2in}
\end{center}
\vspace{.3in}
\baselineskip=24pt

\begin{abstract}
\noindent

A scheme is discussed for embedding 
$n$--dimensional, Riemannian 
manifolds in an $(n+1)$--dimensional 
Einstein space. Criteria for 
embedding a given manifold 
in a spacetime that represents a solution 
to Einstein's equations sourced by 
a massless 
scalar field
are also discussed. 
The embedding procedures are illustrated with a number 
of examples. 

\end{abstract}

\vspace{1.4in}

PACS NUMBERS: 04.50.+h, 04.20.Jb, 98.80.Cq 

\vspace{.3in}

$^1$Electronic address: eda@maths.qmw.ac.uk

$^2$Electronic address: J.E.Lidsey@qmw.ac.uk

\end{titlepage}

\section{Introduction}

\setcounter{equation}{0}

\def\theequation{\thesection.\arabic{equation}}

The possibility that our universe contains hidden, spatial 
dimensions has attracted considerable attention 
over recent years. In particular, 
advances in 
our understanding of the non--perturbative limits of 
superstring theory indicate that spacetime may be 
eleven--dimensional \cite{witten}. A further  
important development has been the realization that 
these extra dimensions 
need not have finite volume. 
Indeed, four--dimensional gravity
can be recovered if the observable 
universe is represented 
by a co--dimension $1$ 
brane embedded in a higher--dimensional space
with a non--factorizable geometry \cite{braneref,hw,lukas,rs1,rs2}. 

Embedding theorems of differential geometry 
provide a natural framework for
relating higher-- and 
lower--dimensional theories of gravity and it 
is important to study such theorems in the light 
of the above developments.  
A particularly powerful theorem, 
due to Campbell, 
states that 
any 
$n$--dimensional, Riemannian manifold can be locally 
and isometrically embedded in an $(n+1)$--dimensional, 
Riemannian space, 
where the Ricci tensor of the latter vanishes \cite{campbell}. 
This theorem was discussed by Romero, Tavakol and 
Zalaletdinov \cite{rtz} within the 
context of the non--compactified approach to Kaluza--Klein 
gravity \cite{ow}. (For earlier work,  see
Refs. \cite{earlier}). Further embeddings into Ricci--flat 
spaces were also established for a wide class of superstring 
backgrounds \cite{lidsey}.  

The purpose of the present paper is to 
develop
extensions of Campbell's scheme, 
where the Ricci tensor 
of the higher--dimensional spacetime
is non--trivial. 
Specifically, we  
consider the case where the
embedding manifold is an Einstein 
space with a 
covariantly constant energy--momentum 
tensor. 
Such embeddings are 
directly 
relevant to the second Randall--Sundrum braneworld 
scenario, where the bulk corresponds to 
pure Einstein gravity sourced only by a 
negative cosmological constant \cite{rs2}. 
The embedding of the four--dimensional,
isotropic and homogeneous radiation universe 
into a Schwarzschild--Anti de Sitter space was recently 
investigated \cite{sv}. Einstein spaces with a positive 
cosmological constant have also become the focus of attention 
\cite{ds,cl} and cosmological solutions 
with such a term 
represent one of the simplest manifestations 
of the inflationary scenario \cite{ll}. 
 
We also consider embeddings into 
spacetimes sourced by a massless, 
minimally coupled scalar field. 
Such a field represents one of the simplest 
forms of matter and can 
be identified, 
after suitable field redefinitions,  
with the dilaton field that arises in the 
string effective action \cite{polch}.
A massless field 
also parametrizes the volume 
of an internal, Ricci--flat space in conventional 
Kaluza--Klein compactification. (For a review, 
see, e.g., \cite{lwc}). 

The paper is organized as follows. 
We develop the embedding schemes in Section 2
and proceed in Section 3 to 
illustrate these techniques by 
establishing  
embeddings for 
general classes of Einstein
and plane wave spacetimes. We conclude with 
a discussion of further 
applications in Section 4.

\section{Embeddings in Higher Dimensions}

\subsection{Einstein Spaces}

\setcounter{equation}{0}

We consider the local 
and isometric embedding of 
the $n$--dimensional, Riemannian manifold, 
$(M, g_{\alpha \beta})$, with 
line element 
\begin{equation}
\label{embeddedmetric}
ds^2 =g_{\alpha\beta} (x^{\mu}) dx^{\alpha}dx^{\beta}
\end{equation} 
in the 
$(n+1)$--dimensional manifold, $(\hat{M}, \hat{h}_{AB} )$,
defined by the metric
\begin{equation}
\label{embeddingmetric}
d\hat{s}^2 =h_{\alpha\beta}dx^{\alpha}dx^{\beta} +\epsilon \phi^2 dy^2
, 
\end{equation}
where $h_{\alpha\beta}=h_{\alpha\beta} (x^{\mu} , y)$ 
and $\phi =\phi (x^{\mu} ,y)$ are analytic functions 
of the $(n+1)$ variables
$\{ x^{\mu} ,y \}$\footnote{Greek and Latin indices run from $(0,1, 
\ldots , n-1)$ and $(0,1, \ldots , n)$, respectively, 
and the coordinate of the 
$(n+1)$th dimension is denoted by $y$. All curvature 
tensors relevant to the $(n+1)$--dimensional 
metric, $\hat{h}_{AB}$, are represented with a  
circumflex accent and those constructed 
from the hypersurface metric, $h_{\alpha\beta}$, 
have no accent. 
We employ Wald's conventions with signature 
$(-, +, +, \ldots )$ for the $n$--dimensional 
spacetime, $g_{\alpha\beta}$ \cite{wald}.
In all cases, 
the embeddings considered 
in this paper are local and 
isometric and do not refer to any aspects of 
the global topology of the spaces.}. 
The constant $\epsilon =\pm 1$ 
and we therefore allow for the possibility that the extra dimension 
is either spacelike or timelike. Although there are well known problems 
with introducing an additional timelike dimension, 
it has been argued that the duality symmetries of 
string/M theory compactified on 
Lorentzian tori result in extra time dimensions
in the strong coupling limit \cite{timehull}. 
Braneworld scenarios where the transverse dimension
is timelike have also been proposed \cite{timebrane}.

When evaluated on an 
arbitrary hypersurface,  
$dy=0$, the components of the Ricci tensor
calculated from the metric (\ref{embeddingmetric}) 
take the general form 
\begin{eqnarray}
\label{ric1}
\hat{R}_{\alpha\beta} =R_{\alpha \beta} -\frac{\nabla_{\alpha \beta}
\phi}{\phi} +\frac{1}{2\epsilon \phi^2} \left( \frac{\phi^*}{\phi} 
h^*_{\alpha\beta} -h^{**}_{\alpha\beta} -\frac{1}{2}h^{\gamma \delta}
h^*_{\gamma \delta} h^*_{\alpha \beta} +h^{\gamma \delta}
h^*_{\alpha \gamma} h^*_{\beta \delta} \right) \\
\label{ric2}
\hat{R}_{\alpha y}= \frac{\phi}{2} \nabla_{\beta} P^{\beta}_{\alpha} \\
\label{ric3}
\hat{R}_{yy}= -\epsilon \phi \nabla^2 \phi -\frac{1}{2} h^{\gamma \delta}
h^{**}_{\gamma \delta} -\frac{1}{2} \left( h^{\gamma\delta} \right)^*
h^*_{\gamma\delta} +\frac{1}{2} h^{\gamma\delta}h^*_{\gamma
\delta} \frac{\phi^*}{\phi} -\frac{1}{4} h^{\gamma\beta}h^{\delta\alpha}
h^*_{\delta \beta} h^*_{\gamma \alpha}  ,
\end{eqnarray}
where the $n$--dimensional Ricci tensor, 
$R_{\alpha\beta}$, and covariant 
derivative operator, $\nabla_{\alpha\beta} = \nabla_{\beta} 
\nabla_{\alpha}$, are 
calculated from $h_{\alpha\beta}$,
a star denotes a partial derivative with respect to $y$, 
$\partial /\partial y |_{y={\rm constant}}$,
evaluated on the hypersurface $y={\rm constant}$, 
$\nabla^2  \equiv h^{\alpha\beta} \nabla_{\alpha\beta}$ and 
the quantity, $P^{\alpha}_{\beta}$, is defined by 
\cite{wesson} 
\begin{equation}
\label{Ptensor}
P^{\beta}_{\alpha} \equiv \frac{1}{\phi} \left( h^{\beta \gamma} 
h^*_{\gamma \alpha} -\delta^{\beta}_{\alpha} h^{\gamma \delta} 
h^*_{\gamma \delta} \right)  .
\end{equation}
If we now define the functions $\Omega_{\alpha\beta}(x^\mu, y)$
\cite{campbell,rtz}:
\begin{equation}
\frac{\partial h_{\alpha\beta}}{\partial y} 
\equiv - 2\phi\Omega_{\alpha\beta} ,
\label{omdef}
\end{equation}
it follows 
that Eqs. (\ref{ric1})--(\ref{ric3}) simplify to
\begin{eqnarray}
\label{ricomg1}
\hat{R}_{\alpha \beta} = R_{\alpha \beta} -
\frac{\nabla_{\alpha \beta} \phi}{\phi} +\epsilon 
\left[ \frac{1}{\phi} \Omega^*_{\alpha \beta} -
h^{\gamma \delta} \Omega_{\gamma\delta} \Omega_{\alpha \beta}
+2 h^{\gamma \delta} \Omega_{\alpha \gamma} 
\Omega_{\beta \delta} \right]  \\
\label{ricomg2}
\hat{R}_{\alpha y} =\phi \nabla_{\beta} 
\left[ \delta^{\beta}_{\alpha} h^{\gamma\delta} \Omega_{\gamma \delta} 
- h^{\beta \gamma} \Omega_{\gamma \alpha} \right]   \\
\label{ricomg3}
\hat{R}_{yy} = - \phi h^{\gamma \beta} \left[ \epsilon 
\nabla_{\gamma \beta} \phi  -\Omega^*_{\gamma \beta} 
-\phi h^{\alpha \delta}\Omega_{\alpha \gamma}
\Omega_{\beta \delta} \right]  .
\end{eqnarray}

In this subsection, we 
show that the
metric (\ref{embeddedmetric}) 
can be embedded in an
Einstein space of the form (\ref{embeddingmetric}), 
where the constraint equations
\begin{equation}
\label{einsteinspace}
\hat{R}_{AB} = \frac{2\Lambda}{1-n} \hat{h}_{AB}
\end{equation}
are satisfied and $\Lambda$ is a spacetime constant.
That 
such an embedding is possible was stated by Campbell
\cite{campbell}, 
but the proof was not given. 
The proof proceeds iteratively by first assuming
that the equations (\ref{einsteinspace}) 
are valid on a specific hypersurface $y=y_0$, 
where $y_0$ is arbitrary, and then 
verifying that they are also valid 
for any $y$ in the neighbourhood 
of this hypersurface. 

To proceed, we substitute Eq. (\ref{einsteinspace}) 
into Eqs. (\ref{ricomg1})--(\ref{ricomg3}): 
\begin{eqnarray}
\label{tensor2}
\Omega_{\alpha\beta}^* =  
h^{\lambda\mu}(\Omega_{\alpha\beta}\Omega_{\lambda\mu} - 
2\Omega_{\alpha\lambda}\Omega_{\beta\mu})\phi + 
\epsilon \nabla_{\alpha\beta} \phi - 
\epsilon\phi R_{\alpha\beta} - 
\frac{2\epsilon\Lambda\phi}{n - 1}h_{\alpha\beta}  \\
\label{precod}
h^{\mu\nu}( \nabla_{\mu}
\Omega_{\alpha\nu} - \nabla_{\alpha} 
\Omega_{\nu\mu}) = 0 \\
\label{scalar2}
h^{\lambda\beta}(\epsilon \nabla_{\lambda\beta} \phi
- \Omega_{\lambda\beta}^* - \phi h^{\alpha\rho}\Omega_{\beta\rho}
\Omega_{\lambda\alpha}) - \frac{2\epsilon\Lambda\phi}{n - 1} = 0  .
\end{eqnarray}
Subtracting the trace of Eq. 
(\ref{tensor2}) from Eq. (\ref{scalar2})
then results in the contracted `Gauss' equation:
\begin{equation}
\label{gauss}
\Omega^2 - \Omega_{\alpha}^{\mu}\Omega_{\mu}^{\alpha} 
= \epsilon(R + 2\Lambda)  ,
\end{equation}
where $\Omega \equiv h^{\alpha\beta}\Omega_{\alpha\beta}$ 
and the covariant constancy of the metric in 
Eq. (\ref{precod}) yields the `Codazzi' equation: 
\begin{equation}
\label{codazzi}
\nabla^{\nu} \Omega_{\alpha \nu} = \nabla_{\alpha} \Omega  .
\end{equation}

A crucial property of the 
higher--dimensional metric (\ref{embeddingmetric}) 
is that it must simplify to the embedded metric 
(\ref{embeddedmetric}) when on the 
hypersurface, $y=y_0$: 
\begin{equation}
\label{restriction}
h_{\alpha\beta}(x^{\mu} ,y_0 ) = g_{\alpha\beta}(x^{\mu}) .
\end{equation}
We then assume that 
the
symmetric functions 
\begin{equation}
\label{symmetric}
\Omega_{\alpha\beta} = 
\Omega_{\beta\alpha}
\end{equation}
can be found that 
satisfy the constraints (\ref{gauss}) and (\ref{codazzi}) 
on this `initial'  hypersurface. Moreover, 
it is also assumed that these functions 
evolve according to Eq. (\ref{omdef}) 
and the set of differential equations\footnote{The metric $h_{\alpha\beta} (
x^{\mu}, y)$ is employed in Eq. (\ref{omegadevelop})
to raise and lower indices and in 
calculating the curvature tensors and 
covariant derivatives.}
\begin{equation}
\label{omegadevelop}
\frac{\partial \Omega^{\gamma}_{\beta}}{\partial y} = 
\epsilon \nabla^{\gamma}_{\beta}  \phi + \phi 
\left( \Omega \Omega^{\gamma}_{\beta} -\epsilon R^{\gamma}_{\beta} 
-\frac{2\epsilon \Lambda}{n-1} \delta^{\gamma}_{\beta} \right)   ,
\end{equation}
where the boundary conditions
\begin{equation}
\label{bc}
h^*_{\alpha\beta} =-2\phi(x^{\mu} , y_0) \Omega_{\alpha\beta} 
(x^{\mu} ,y_0)
\end{equation}
are satisfied. 

As shown in the appendix, 
if conditions (\ref{omdef}), (\ref{gauss})--(\ref{bc})
are satisfied, it follows that 
\begin{eqnarray}
\label{deriv1}
\left( \nabla^{\nu} \Omega_{\alpha \nu}
-\nabla_{\alpha} \Omega \right)^* =0 \\
\label{deriv2}
\left( \Omega^2 
-\Omega^{\beta}_{\alpha} \Omega^{\alpha}_{\beta} -\epsilon 
(R +2\Lambda ) \right)^* =0
\end{eqnarray}
and Eqs. (\ref{deriv1}) and (\ref{deriv2}) 
then imply that 
Eqs. (\ref{gauss}), (\ref{codazzi}) and 
(\ref{symmetric}) are valid for all 
hypersurfaces, $dy=0$, in the {\em neighbourhood} of 
$y=y_0$. 
Given the validity of Eq. (\ref{omegadevelop}), 
therefore, we may further 
deduce that 
the Einstein conditions	
(\ref{einsteinspace}) are satisfied  
for all $y$ in this neighbourhood. 
Consequently, 
the $( \alpha \beta )$--components 
of the higher--dimensional metric $\hat{h}_{AB}$ can be 
expanded as a Taylor series in $y$ to first--order:
\begin{equation}
\label{firstorder}
\hat{h}_{\alpha\beta} = g_{\alpha\beta} -
2\phi (x^{\mu} ,y_0 ) \Omega_{\alpha\beta} (
x^{\mu} ,y_0) y  ,
\end{equation}
where Eqs. (\ref{omdef}) and (\ref{restriction}) 
have been employed. Likewise, 
the value of $\Omega_{\alpha\beta}$ 
in this vicinity can be determined 
from 
Eq. (\ref{omegadevelop}). 
Since the analysis 
is valid for an arbitrary hypersurface, 
this local extension can be repeated 
recursively and this establishes the 
embedding 
of the metric (\ref{embeddedmetric}) 
in the Einstein space (\ref{embeddingmetric}). 
The embedding is not unique since 
more than one choice of 
$\Omega_{\alpha\beta}$ may be possible for a given 
embedded metric 
\cite{lidsey}.
When $n=3$ and $\epsilon =-1$, 
the determination of $\Omega_{\alpha\beta}$ 
from the point of view of the boundary value 
problem follows by identifying 
$y$ with the timelike coordinate, $h_{\alpha\beta}$ 
with the spatial three--metric and 
$\Omega_{\alpha\beta}$ with the extrinsic curvature
\cite{wald}. 
The conditions that these
functions must satisfy are the five 
relations (\ref{gauss})--(\ref{codazzi}) and 
(\ref{symmetric})--(\ref{bc}). 
We may conclude, 
therefore,
that 
any $n$--dimensional, 
Riemannian manifold may be locally and 
isometrically embedded in an $(n+1)$--dimensional 
Einstein space when Eqs. (\ref{gauss})--(\ref{bc}) 
are satisfied. 
 
\subsection{Massless Scalar Fields}

It is also of interest to consider 
whether Campbell's technique 
can be extended to 
include embeddings of the metric (\ref{embeddedmetric}) 
in non--vacuum  
spacetimes, $( \hat{M} , \hat{h}_{AB} )$. 
One possible source of matter is a 
massless, minimally coupled scalar field, $\chi$, 
that satisfies the Einstein 
field equations
\begin{eqnarray}
\label{scalarfield1}
\hat{R}_{AB} =\frac{1}{2} \hat{\nabla}_A \chi \hat{\nabla}_B
\chi \\
\label{scalarfield2}
\hat{h}^{AB} \hat{\nabla}_{AB} \chi =0   .
\end{eqnarray}

For the metric ansatz (\ref{embeddingmetric})
the components of the 
$(n+1)$--dimensional Ricci tensor 
(\ref{scalarfield1}) reduce 
to 
\begin{eqnarray}
\label{einsteinscalar1}
R_{\alpha \beta} -
\frac{\nabla_{\alpha \beta} \phi}{\phi} 
+ \epsilon \left[  \frac{1}{\phi} \Omega^*_{\alpha \beta} -
h^{\gamma \delta} \Omega_{\gamma\delta} \Omega_{\alpha \beta}
+2 h^{\gamma \delta} \Omega_{\alpha \gamma} 
\Omega_{\beta \delta} \right]  
= \frac{1}{2} \nabla_{\alpha} \chi \nabla_{\beta} \chi
\\
\label{einsteinscalar2}
\phi
\nabla_{\beta} 
\left[ \delta^{\beta}_{\alpha} h^{\gamma\delta} \Omega_{\gamma \delta} 
- h^{\beta \gamma} \Omega_{\gamma \alpha} \right]  
= \frac{1}{2} \chi^* \nabla_{\alpha} \chi \\
\label{einsteinscalar3}
\phi h^{\gamma \beta} \left[ 
\epsilon \nabla_{\gamma \beta} \phi  -\Omega^*_{\gamma \beta} 
-\phi h^{\alpha \delta}\Omega_{\alpha \gamma}
\Omega_{\beta \delta} \right]  = 
- \frac{1}{2} \left( \chi^* \right)^2
\end{eqnarray}
on the hypersurface, $y=y_0$, 
where the symmetric functions, 
$\Omega_{\alpha\beta}=\Omega_{\beta\alpha}$, 
are defined, as before, in Eq. (\ref{omdef})
and the 
boundary conditions (\ref{restriction}) and (\ref{bc}) are 
also assumed to be valid. 
The left--hand side of the 
scalar field equation (\ref{scalarfield2}) 
takes the form
\begin{equation}
\label{einsteinscalar4}
h^{\alpha\beta} 
\nabla_{\alpha\beta} \chi +\frac{1}{\phi} h^{\alpha\beta}
\nabla_{\alpha} \phi \nabla_{\beta} \chi +
\frac{\epsilon}{\phi} \left[ 
\frac{1}{\phi} \chi^{**} - \frac{1}{\phi^2} 
\phi^* \chi^* - \Omega \chi^* \right] =0
\end{equation}
on this  hypersurface
and taking the trace of Eq. (\ref{einsteinscalar1})
and subtracting Eq. (\ref{einsteinscalar3}) implies that 
\begin{equation} 
\label{einsteinscalar5}
\Omega_{\alpha\beta} \Omega^{\alpha\beta}
-\Omega^2 = \epsilon \left[ 
\frac{1}{2} \nabla_{\alpha}\chi \nabla^{\alpha} \chi
-R -\frac{\epsilon}{2\phi^2} \left( \chi^*
\right)^2 \right]   .
\end{equation}

We now consider 
the scalar function, $\chi$,
and 
the symmetric functions, 
$\Omega_{\alpha\beta}$,  
that satisfy
the conditions (\ref{einsteinscalar2}), 
(\ref{einsteinscalar4}) and (\ref{einsteinscalar5}) on the hypersurface, 
$y=y_0$, and
evolve according to 
\begin{equation}
\label{omegascalar}
\frac{\partial \Omega^{\gamma}_{\beta}}{\partial y} 
= \epsilon \nabla^{\gamma}_{\beta}  \phi - 
\epsilon \phi \left[ 
R^{\gamma}_{\beta} -\frac{1}{2} \nabla^{\gamma} \chi 
\nabla_{\beta} \chi \right] +\Omega \Omega^{\gamma}_{\beta} 
\phi  .
\end{equation}
It can then be shown, 
following 
an argument similar to that presented in the Appendix, 
that 
\begin{eqnarray}
\label{ontheway}
\left( \nabla_{\lambda} \Omega^{\lambda}_{\mu} \right)^*
= \Omega^{\eta}_{\kappa} \left[ 
\Omega^{\kappa}_{\eta} \nabla_{\mu} \phi +\phi 
\nabla_{\mu} \Omega^{\kappa}_{\eta} \right] - \Omega^{\rho}_{\mu}\left[ 
\Omega \nabla_{\rho} \phi +\phi \nabla_{\rho} \Omega \right] 
\nonumber \\
+\epsilon \nabla_{\mu} \nabla^2 \phi +\Omega \Omega^{\lambda}_{\mu} 
\nabla_{\lambda} \phi +\frac{\epsilon}{2} \nabla_{\lambda} 
\phi \nabla_{\mu} \chi \nabla^{\lambda} \chi
\nonumber \\
+\phi \left[ \Omega^{\lambda}_{\mu} 
\nabla_{\lambda} \Omega 
-\epsilon \nabla_{\lambda} R^{\lambda}_{\mu} 
+\Omega \nabla_{\lambda} \Omega^{\lambda}_{\mu} +
\frac{\epsilon}{2} \nabla^{\lambda} \chi \nabla_{\mu\lambda} \chi
+\frac{\epsilon}{2} \nabla_{\mu} \chi  \nabla^2 \chi \right]  ,
\end{eqnarray}
where we have employed Eq. (\ref{omegascalar}), 
(\ref{Riccilemma}) 
and (\ref{Gammaderiv1})--(\ref{Gammaderiv2}).
Evaluating the derivative with respect to $x^{\mu}$ 
of the trace of Eq. (\ref{omegascalar}) and combining 
the result with 
Eq. (\ref{ontheway}) then 
implies that 
\begin{equation}
\label{star1}
\left. \left( \nabla_{\lambda} \Omega^{\lambda}_{\mu} 
-\nabla_{\mu} \Omega +\frac{1}{2\phi} \chi^* 
\nabla_{\mu} \chi
\right)^* =\frac{\epsilon}{2} \phi \nabla_{\mu} 
\chi \left( 
\hat{h}^{AB} \hat{\nabla}_{AB} \chi
\right) \right|_{y=y_0}   ,
\end{equation}
where 
we have substituted for the scalar 
field equation (\ref{einsteinscalar4}), 
employed Eqs. (\ref{einsteinscalar2}) and 
Eqs. (\ref{einsteinscalar5})
and the right--hand side is evaluated at 
$y=y_0$. 
It follows from Eq. (\ref{einsteinscalar4}), 
therefore, 
that 
the right--hand side of 
Eq. (\ref{star1}) vanishes. 
Consequently, 
Eq. (\ref{einsteinscalar2}) is 
also valid 
for all hypersurfaces in the neighbourhood of $y=y_0$.

The validity of Eq. (\ref{einsteinscalar5}) in 
this neighbourhood
is established 
by determining the derivative of the Ricci scalar, 
$R(h)$, with 
respect to the variable, $y$. We find that 
\begin{eqnarray}
\label{Rstarchi}
R^* = 2\phi \Omega^{\mu}_{\beta} R^{\beta}_{\mu} 
+2\Omega \nabla^2 \phi -2\Omega^{\mu}_{\tau} \nabla^{\tau}_{\mu} 
\phi \nonumber \\
+\chi^* \nabla^2 \chi +\nabla^{\mu} \chi \nabla_{\mu} 
\left( \chi^* \right) 
+ \frac{1}{\phi} \chi^* \nabla_{\mu} \chi \nabla^{\mu} \phi .
\end{eqnarray}
Moreover, it follows from Eq. (\ref{omegascalar}) 
and its trace that
\begin{eqnarray}
\label{omegastarchi}
\left( \Omega^{\mu}_{\lambda} \Omega^{\lambda}_{\mu} 
-\Omega^2 \right)^* = 2\epsilon \Omega^{\beta}_{\alpha} 
\nabla^{\alpha}_{\beta} \phi -2\epsilon \Omega \nabla^2 
\phi -2 \epsilon 
\phi \Omega^{\beta}_{\alpha} R^{\alpha}_{\beta} 
\nonumber \\
+\epsilon \phi \Omega^{\beta}_{\alpha} \nabla_{\beta} \chi 
\nabla^{\alpha} \chi -\frac{\Omega}{\phi} \left( \chi^* 
\right)^2
\end{eqnarray}
and combining Eqs. (\ref{Rstarchi}) and 
(\ref{omegastarchi}) 
implies that 
\begin{equation}
\label{star2}
\left. \left( \Omega_{\alpha\beta} \Omega^{\alpha\beta} 
-\Omega^2 +\epsilon R -\frac{\epsilon}{2} \nabla^{\alpha} \chi 
\nabla_{\alpha} \chi + \frac{1}{2\phi^2}
\left( \chi^* \right)^2 \right)^*
= \epsilon \chi^* \hat{h}^{AB} \hat{\nabla}_{AB} 
\chi \right|_{y=y_0} .
\end{equation}

The right--hand side of Eq. (\ref{star2}) 
also 
vanishes when Eq. (\ref{einsteinscalar4}) 
is satisfied and this implies that Eq. 
(\ref{einsteinscalar5}) is valid in the neighbourhood 
near $y=y_0$. 
We may conclude, therefore, 
that the field equations (\ref{scalarfield1}) are 
valid for all $y$ and 
Eq. (\ref{scalarfield2}) then follows from the contracted 
Bianchi identity. 

To summarize, 
for the functions 
$\Omega_{\alpha\beta} = 
\Omega_{\alpha\beta} (x^{\mu}, y)$, 
$\phi =\phi (x^{\mu}, y)$ and 
$\chi = 
\chi (x^{\mu}, y )$ 
that 
satisfy
Eqs. (\ref{einsteinscalar2}), (\ref{einsteinscalar4}) 
and (\ref{einsteinscalar5}) on the hypersurface $y=y_0$, 
and evolve according 
to Eqs. (\ref{omdef}), 
(\ref{restriction}), 
(\ref{bc}) 
and (\ref{omegascalar}), 
the metric (\ref{embeddedmetric}) can be 
embedded in the manifold (\ref{embeddingmetric}),
where the  
latter is a 
solution to the Einstein field equations 
(\ref{scalarfield1}) and (\ref{scalarfield2}) for a massless, 
minimally coupled scalar field.

This concludes our discussion of the 
embedding schemes. In the following Section, 
we employ the procedures 
to embed 
Einstein and plane wave spacetimes 
in non--vacuum, higher--dimensional manifolds. 

\section{Applications of the Embedding Schemes}

\setcounter{equation}{0}

\def\theequation{\thesection.\arabic{equation}}

\subsection{Embedding Einstein Spaces in Einstein Spaces}

We first consider the embedding of an 
$n$--dimensional 
Einstein space
\begin{equation} 
\label{neinstein} 
R_{\alpha\beta} (g) =\frac{2\lambda}{2-n} g_{\alpha\beta}
\end{equation}
in the 
$(n+1)$--dimensional Einstein space (\ref{einsteinspace}) 
for arbitrary constants $\{ \lambda , \Lambda \}$. 
The embedding is achieved by invoking 
the {\em ansatz}
\begin{equation}
\label{ansatz}
\Omega_{\alpha \beta} \equiv C h_{\alpha \beta} , \qquad \phi =1  ,
\end{equation}
where $C=C(x^{\mu} , y)$ is a scalar function. 
Eq. (\ref{codazzi}) immediately implies that 
the prefactor, $C$, 
may be a function only of $y$. On the other
hand, 
Eq. (\ref{gauss}) implies that 
\begin{equation}
\label{Cconstraint}
C^2 = \frac{\epsilon (R+2\Lambda )}{n(n-1)}
\end{equation} 
and, 
since
$\phi=1$, it follows that Eq. (\ref{omdef}) 
may be formally integrated to yield
\begin{equation}
\label{hexpression}
h_{\alpha\beta} = a^2 (y) g_{\alpha\beta} ,
\end{equation}
where 
the `warp factor', $a$, is defined 
by  $a \equiv \exp \left[ - \int^y dy' C(y') \right]$.
If the constant of integration is chosen such that 
$a(y_0 ) =1$, the 
$n$--dimensional 
metric $g_{\alpha\beta}$ may be interpreted as the 
embedded Einstein space satisfying 
Eq. (\ref{neinstein}). Indeed, 
only Eqs. (\ref{omegadevelop}) and (\ref{Cconstraint})
remain to be solved for the 
embedding to be determined and these equations reduce to 
\begin{eqnarray}
\frac{1}{a} \frac{d^2 a}{dy^2} 
=\frac{2\epsilon \Lambda}{n(n-1)} \\
\left(  
\frac{da}{dy} \right)^2 = 
\frac{2\epsilon \Lambda}{n(n-1)} a^2 -
\frac{2\epsilon \lambda}{(n-1)(n-2)}  ,
\end{eqnarray}
respectively.  
The general solution satisfying the boundary 
condition (\ref{bc}) 
is then given by 
\begin{equation}
\label{aexpression}
a = {\rm cosh} \left( \sqrt{\frac{2\epsilon \Lambda}{n(n-1)}} 
(y -y_0) 
\right) +B {\rm sinh} \left( \sqrt{\frac{2\epsilon \Lambda}{n(n-1)}} 
(y -y_0) 
\right)   ,
\end{equation}
where 
\begin{equation}
\label{Bexpression}
B^2  = 1-\frac{n\lambda}{(n-2)\Lambda}
\end{equation}
and the embedding of the Einstein space (\ref{neinstein}) 
is therefore given by 
\begin{equation}
\label{nonlinear}
d\hat{s}^2 = a^2 (y) g_{\alpha\beta} dx^{\alpha}
dx^{\beta} + dy^2   ,
\end{equation}
where 
Eqs.  (\ref{aexpression})
and (\ref{Bexpression}) are satisfied. 

Eq. (\ref{nonlinear}) generalizes 
the embedding of maximally symmetric, 
four--dimensional Einstein spaces 
in five dimensions \cite{kn} as well as the 
embedding found in Ref. \cite{gs}
for $\lambda < 0$. When the 
embedded manifold is Ricci--flat $(\lambda =0)$, 
the warp factor (\ref{aexpression}) 
is exponential and Eq. (\ref{nonlinear}) reduces to the 
metric considered in Ref. \cite{clp}.

One interesting 
consequence of the embedding 
(\ref{nonlinear}) is that 
it provides the bulk solution 
for non--fine--tuned versions 
of the Randall--Sundrum--type braneworld 
scenarios, where the co--dimension $1$ branes 
have a non--vanishing cosmological constant 
\cite{kn,kr,big,sabra}. Since 
the embedded metric is arbitrary in our 
analysis, it may be viewed as a non--linear 
generalization of the 
graviton zero mode on the brane. 
Within this context, a specific 
example 
is given by 
the Siklos class of solutions to 
Eq. (\ref{neinstein}) 
representing 
gravitational waves propagating 
in anti--de Sitter spacetime \cite{siklos,gc}. 

\subsection{Plane Waves}

We now consider 
the embedding 
of 
the plane wave backgrounds \cite{brinkmann}
\begin{equation}
\label{wavemetric}
ds^2 =-dudv + du^2 +f_{ij} dx^i dx^j
\end{equation}
in a manifold sourced by a massless 
scalar field, $\chi$, 
following the approach outlined in Section 2.2
for $\epsilon =1$. 
The arbitrary function 
$f_{ij}=f_{ij}(u)$ is symmetric
and depends only on the light--cone coordinate, $u$.
The metric (\ref{wavemetric}) 
admits a covariantly constant, null Killing 
vector field, 
$\partial / \partial v$, that is orthogonal 
to the Riemann curvature tensor. 
Consequently, all 
curvature invariants vanish and this implies that 
metrics of this form 
can represent perturbatively 
exact solutions to the string equations 
of motion when the dilaton 
and 
antisymmetric form fields satisfy appropriate 
conditions \cite{hs,tset}. The only non--trivial component 
of the Ricci tensor is $R_{uu}$ 
and is also 
a function only of $u$.

To proceed with the embedding, we assume 
that the scalar field is independent 
of the coordinate, $y$, 
and further 
invoke 
the {\em ansatz}
\begin{eqnarray}
\label{omegaansatz}
\Omega_{\alpha\beta} =\left\{ 
\begin{array}{ll} 
y /y_0^2 & \mbox{if $\alpha = \beta =u$} \\    
0 & \mbox{otherwise} 
\end{array}
\right. 
\end{eqnarray}
together with the condition  
\begin{equation}
\label{phiant}
\phi =-1  .
\end{equation}

On the hypersurface $y=y_0$, 
where indices are raised 
with $g^{\alpha\beta}$, the only non--trivial 
components of $\Omega^{\alpha}_{\beta}$ and 
$\Omega^{\alpha\beta}$ are $\Omega^{vv} = 2\Omega^v_u
=4\Omega_{uu}$. Thus, Eq. (\ref{einsteinscalar2}) 
is solved since 
the embedded metric (\ref{wavemetric}) and 
$\Omega^{\alpha}_{\beta}$ are both independent of 
$v$. Eq. (\ref{einsteinscalar5}) is also satisfied 
if the scalar field is a function only of 
$u$, $\chi =\chi (u)$, and this 
latter condition also ensures that 
Eq. (\ref{einsteinscalar4}) holds when $y=y_0$. 
We may then solve the set of 
equations (\ref{omdef}) to deduce that 
\begin{eqnarray}
\label{hsol}
h_{\alpha \beta} =\left\{ 
\begin{array}{ll}
\left( y /y_0 \right)^2 & \mbox{if $\alpha = \beta =u$} \\
g_{\alpha\beta} & \mbox{otherwise}
\end{array} \right. 
\end{eqnarray}
This implies that in the neighbourhood of the hypersurface, 
only $\Omega^{v}_u$ and $\Omega^{vv}$ are non--trivial and, 
consequently, Eqs. (\ref{einsteinscalar2}) and 
(\ref{einsteinscalar5}) are solved for arbitrary $y$.
Moreover, Eq. (\ref{scalarfield2}) is trivially satisfied, 
since the scalar field is null. 
Thus, only Eq. (\ref{omegascalar}) is yet to be solved 
and this set of conditions 
reduces to 
the single 
constraint: 
\begin{equation}
\label{singexp}
\left( \frac{d \chi}{d u} \right)^2 =2 \left[ R_{uu} -\frac{1}{y_0^2} 
\right]    ,
\end{equation}
where $R_{uu}$ is the 
$(uu)$--component of the Ricci tensor calculated 
from the embedded metric (\ref{wavemetric}). 
We may conclude, therefore, 
that  
the $(n+1)$--dimensional embedding metric 
is given by 
\begin{equation}
\label{hatwave}
d\hat{s}^2 = -dudv + \left( \frac{y}{y_0} \right)^2 du^2 
+f_{ij}dx^idx^j +dy^2  ,
\end{equation}
where the scalar field is determined by the quadrature 
\begin{equation}
\label{singlequadrature}
\chi = \sqrt{2} \int^u du' \left[ R_{uu} (u') -\frac{1}{y_0^2} 
\right]^{1/2}  .
\end{equation}

An interesting example of this embedding arises 
for the four--dimensional backgrounds
defined by 
$f_{ij} = f^2(u) \delta_{ij}$, where 
$\delta_{ij}$ is the two--dimensional Kronecker delta
and 
$f=f(u)$ is an arbitrary function 
that parametrizes 
the amplitude of the plane wave.  
The Ricci tensor for such a metric is given by $R_{uu}=-2f^{-1}
(d^2f/du^2)$ and Eq. (\ref{singexp}) therefore 
has the form of a 
one--dimensional Helmholtz equation: 
\begin{equation}
\label{helmholtz}
\left[  \frac{d^2}{du^2} +V(u) \right] f =0  ,
\end{equation}
where the effective potential, $V(u)$, is determined 
by the kinetic energy of the 
scalar field: 
\begin{equation}
\label{potential}
V \equiv \frac{1}{2} \left[ \frac{1}{2} 
\left( \frac{d \chi}{d u} \right)^2 + \frac{1}{y_0^2} \right] .
\end{equation}
It follows that if a particular solution, 
$f_1 (u)$, 
to Eq. (\ref{helmholtz}) 
can be found for a given choice of $\chi (u)$, 
the general solution can be expressed directly in 
terms of this solution such that 
\begin{equation}
\label{direct}
f_{\rm gen}= \left[ \kappa + \int^{u} \frac{du'}{f^2_1(u')}
\right] f_1 (u) ,
\end{equation}
where $\kappa$ is an arbitrary constant. 
In general, this implies that 
there is not a one--to--one 
correspondence between the amplitude of the 
embedded metric and the functional 
form of the scalar 
field that generates the Ricci curvature 
of the embedding metric. 

Finally, a second metric of interest is the Nappi--Witten 
WZW model 
\begin{equation}
ds^2 =-dudv +du^2 +dx^2 +2 \cos u dx dy +dy^2
\end{equation}
that corresponds to a conformal field theory 
describing a homogeneous, monochromatic plane wave \cite{nw}. 
The Ricci tensor of this background is 
$R_{uu}=1/2$, implying that the scalar 
field takes the particularly simple 
form $\chi = [1-(2/y^2_0)]^{1/2} u$.

\subsection{Embeddings of Ricci--flat Spaces}

We conclude this Section by establishing a class of embeddings
where the 
scalar field is independent of the coordinates 
of the embedded metric 
(\ref{embeddedmetric}), i.e., 
$\chi$ is a function only of $y$. We 
employ the {\em ansatz} (\ref{ansatz}) 
for the functions $\{ \Omega_{\alpha\beta} ,\phi \}$. In this 
case, Eq. (\ref{einsteinscalar2}) implies that 
$\nabla_{\alpha} C =0$ and Eq. (\ref{einsteinscalar5})
implies that 
\begin{equation}
\label{gaussC}
n(n-1) C^2 =\epsilon R +\frac{1}{2} \left( 
\frac{d \chi}{dy} \right)^2  .
\end{equation}
The scalar field equation (\ref{einsteinscalar4}) simplifies to 
\begin{equation}
\label{scalarC}
\frac{d^2 \chi}{d y^2} = nC \frac{d \chi}{dy}
\end{equation}
and the evolution equation (\ref{omegascalar}) takes the form
\begin{equation}
\label{ricexp}
R^{\gamma}_{\beta} =-\epsilon \left( \frac{dC}{dy} -n
C^2 \right) \delta^{\gamma}_{\beta}  .
\end{equation}

Eq. (\ref{scalarC}) 
admits the first integral 
\begin{equation}
\label{firstC}
\left( \frac{d\chi}{dy} \right)^2 =\frac{m^2}{a^{2n}}  ,
\end{equation}
where $m$ is an arbitrary constant and 
we have defined the function 
$C \equiv - d \ln a /dy$. Taking the trace of 
(\ref{ricexp}) and combining Eqs. (\ref{gaussC}) 
and (\ref{firstC}) 
then implies that 
\begin{equation}
\label{aequationC}
\frac{d^2 a}{dy^2} =-\frac{m^2}{2n} a^{1-2n}
\end{equation}
and Eq. (\ref{aequationC}) may be integrated
to yield the solution
\begin{equation}
\label{asolutionC}
a = [1+b(y-y_0)]^{1/n}   ,
\end{equation}
where $b^2 \equiv nm^2/[2(n-1)]$. 
By substituting Eq. (\ref{asolutionC}) 
into Eq. (\ref{ricexp}), 
we then deduce that the Ricci tensor 
of the embedded metric must vanish. Finally, 
the form of the embedded metric follows 
by integrating 
Eq. (\ref{omdef}):
\begin{equation}
\label{flathigher}
d\hat{s}^2 = [1+b(y-y_0)]^{2/n} g_{\mu\nu} dx^{\mu} dx^{\nu}
+\epsilon dy^2
\end{equation}
and this satisfies the boundary conditions 
(\ref{restriction}) and (\ref{bc}). 

Thus, Eq. (\ref{flathigher}) represents the 
the embedding of an arbitrary, 
$n$--dimensional, 
Ricci--flat spacetime into a manifold sourced 
by a massless scalar field, 
where the latter is given by 
$\chi 
=(m/b) \ln [1+b(y-y_0)]$. 
In the special case where the embedded metric 
is four--dimensional, flat Minkowski space, Eq. (\ref{flathigher}) 
represents the bulk metric for 
a class of braneworld models, where the 
cosmological constant on the brane is arbitrary but 
has no influence on the brane dynamics
\cite{selftune}. 
These models are interesting 
because they 
provide a
new perspective on 
solving the 
cosmological constant problem. The 
embedding considered in this subsection 
indicates that 
this problem may also discussed 
within a wider context \cite{feinstein}. 

\section{Discussion}

In this paper, 
we have developed a procedure,
introduced by Campbell,  
to embed a given Riemannian 
manifold into an Einstein space
with a non--trivial cosmological constant. 
Such 
an embedding has a number of applications. 

Firstly, the scheme is iterative and does not 
depend on the dimensionality of the embedded 
space. Thus, if the embedding 
of a particular $n$--dimensional space, 
$M$, in an $(n+1)$--dimensional 
Einstein space, $M_{\rm Ein}$, can be determined, 
an embedding of the space $M$ into  
an $(n+2)$--dimensional Einstein space follows immediately 
by embedding $M_{\rm Ein}$ along the lines outlined in 
Section 3.1. 

This provides a method for 
generating and classifying exact solutions 
to higher--dimensional theories of gravity. 
For example, the infra--red limit of M--theory 
is eleven--dimensional supergravity, 
with a bosonic 
sector consisting of 
the graviton and a three--form 
antisymmetric potential \cite{witten}. 
Recently, 
it was shown that the field equations 
for this theory can be written in such a way 
that only ten--dimensional Poincare invariance is 
manifest \cite{hlw}. This is equivalent to 
performing a 
generalized Scherk--Schwarz dimensional 
reduction to ten dimensions, where the fields 
are allowed to depend 
specifically 
on the 
compactifying 
coordinate \cite{schsch}. The resulting 
ten--dimensional theory 
represents a `massive'
extension of type IIA supergravity.  
It was further 
shown that if the gauge fields are then frozen out, 
the ten--dimensional equations of motion reduce to the 
single equation \cite{hlw}
\begin{equation}
\label{massiveIIA}
\hat{R}_{AB} = m^2 \hat{h}_{AB}  ,
\end{equation}
where 
$m^2$ represents a cosmological 
constant. Thus, the embeddings 
that we have discussed in this paper 
may be employed to 
generate solutions 
to the massive type IIA supergravity and 
eleven--dimensional supergravity theories. 

Embeddings in Einstein spaces are also relevant to 
Wesson's `spacetime--matter' (STM) theory, 
where the matter on 
any $(3+1)$--dimensional hypersurface is
encoded at a classical level purely in terms of
five-dimensional vacuum geometries \cite{ow,wessonbook}.
As discussed in Ref. \cite{rtz}, this interpretation is closely linked
to that of Campbell's theorem \cite{campbell}.  Thus, embeddings
in Einstein spaces would be related 
to a generalisation of the STM theory,
although such a generalization 
could only be achieved 
at the price of introducing a curvature 
length scale. 
It would be of interest to investigate the relationship between
four--dimensional matter and the geometry of 
five--dimensional Einstein spaces further.
Moreover, such a generalization 
would enable direct comparisons
to be made 
between the STM theory and braneworld models. 
In particular, both approaches 
attempt to 
attach physical significance to the fifth coordinate
\cite{wessonbook,bk} and 
these attempts should share some common obstacles and 
insights.

In establishing the embedding of Einstein spaces 
we invoked the ansatz (\ref{ansatz}). 
This restriction could be relaxed by allowing 
$\Omega_{\alpha\beta}$ to have more degrees 
of freedom. One possibility 
is to specify 
$\Omega_{\alpha\beta} = Q_{\alpha}^{\gamma}h_{\gamma\beta}$, 
where $Q^{\alpha}_{\gamma}$ has the block--diagonal form 
\begin{equation}
Q^{\alpha}_{\gamma} = {\rm diag} 
\left[ C(x^A), \ldots, C(x^A), D(x^A) ,  ... , D(x^A)
\right]  
\end{equation}
for some scalar functions $\{ C ,D \}$. 
It would be natural to consider such 
an ansatz 
when embedding an Einstein space that 
itself is the product of two or more lower--dimensional 
Einstein spaces. 
An alternative approach -- relevant 
to spatially homogeneous cosmologies --
is to first embed the $(n-1)$--dimensional
spacelike hypersurface 
in a space with an extra spatial dimension 
and to then view the embedding to $(n+1)$ 
dimensions as 
an initial
value problem \cite{wald}.

The embedding of manifolds in Einstein gravity 
with a massless scalar field can 
also
provide the seed for generating new, 
higher--dimensional solutions 
to the string equations of motion. 
In the case
where the embedded metric admits an Abelian 
isometry associated with a Killing vector, 
$\partial /\partial z$,
a conformal transformation 
to the string frame, followed
by a T--duality 
transformation, may be 
performed. This symmetry transformation 
inverts the 
string--frame metric coefficient 
associated with $z$ and results in a new dilaton
field. Solutions with non--trivial 
form fields may also be found by employing further 
duality transformations \cite{lwc,tset}.

Finally, we remark that since the 
cosmological constant and scalar 
field considered in Section 2 were uncoupled, 
it follows that 
embeddings in manifolds sourced by both 
degrees of freedom can in principle be found
by extending the above analyses. In particular, we 
may deduce immediately that  
if a solution, 
$\{ g_{\alpha\beta} , \chi \}$, to the 
$n$--dimensional field equations 
\begin{eqnarray}
\label{masslesslambda1}
R_{\alpha\beta} (g) = \frac{1}{2} 
\nabla_{\alpha} \chi \nabla_{\beta} \chi 
+ \frac{2\lambda}{2-n} g_{\alpha\beta} \\
\label{masslesslambda2}
g^{\alpha\beta} \nabla_{\alpha\beta} \chi =0
\end{eqnarray}
is known, the 
metric 
\begin{equation}
\label{metricplus}
d\hat{s}^2 =a^2(y) g_{\alpha\beta} dx^{\alpha} dx^{\beta} 
+dy^2
\end{equation}
represents a solution 
of the equations of motion derived from 
the $(n+1)$--dimensional 
action
\begin{equation}
\label{action1}
S=\int d^{n+1} x \sqrt{-\hat{h}} \left[ 
\hat{R}-\frac{1}{2} \left( 
\hat{\nabla} \chi \right)^2 - 2 \Lambda \right] ,
\end{equation}
where the warp factor, $a=a(y)$, is given by 
Eqs. (\ref{aexpression}) and (\ref{Bexpression}),
and the functional form of the scalar field is 
unaltered. Scalar field spacetimes satisfying 
Eqs. (\ref{masslesslambda1}) and 
(\ref{masslesslambda2}) were recently 
studied within the context of the AdS/CFT correspondence 
\cite{adscft}. 
This embedding generalizes 
the embedding for $\lambda =0$ found 
recently by Feinstein, Kunze and Vazquez--Mozo
\cite{feinstein}.  

\vspace{.7in}

{\bf Acknowledgements} EA is supported by 
the Particle Physics and Astronomy Research 
Council (PPARC) and JEL 
by the Royal Society. We thank J. Barbour and 
R. Tavakol for discussions.

\appendix

\section{Appendix}

\setcounter{equation}{0}

\def\theequation{A.\arabic{equation}}

In this appendix we derive the conditions (\ref{deriv1}) and 
(\ref{deriv2}). In doing so, we employ 
the expressions 
\begin{eqnarray}
\label{Riccilemma}
\nabla_{\mu} \nabla^2 \phi =
\nabla_{\lambda} \left( \nabla^{\lambda}_{\mu} \phi \right) 
-R^{\lambda}_{\mu} 
\nabla_{\lambda} \phi \\
\label{rielem}
\left( \nabla_{\beta\alpha} -\nabla_{\alpha\beta} 
\right) T^{\gamma}_{\delta} =- 
{R_{\alpha\beta \epsilon}}^{\gamma} T^{\epsilon}_{\delta}
+{R_{\alpha\beta\delta}}^{\epsilon} T^{\gamma}_{\epsilon}
\end{eqnarray}
for a scalar field and 
a tensor field $T^{\gamma}_{\delta}$
derived from the Ricci lemma \cite{wald}. 
We also require expressions  
for the $y$--derivative of the Christoffel matrices 
\cite{campbell}: 
\begin{eqnarray} 
\label{Gammaderiv1}
\Omega^{\kappa}_{\eta} 
\left( \Gamma^{\eta}_{\rho\kappa}
\right)^* =- \left( \phi \nabla_{\rho} \Omega^{\eta}_{\kappa} + 
\Omega^{\eta}_{\kappa} \nabla_{\rho} \phi \right) 
\Omega^{\kappa}_{\eta} 
\\
\label{Gammaderiv2}
\left( \Gamma^{\kappa}_{\eta\kappa} \right)^* = 
- \nabla_{\eta} \left( \phi \Omega \right) .
\end{eqnarray}

We first consider Eq. (\ref{deriv1}). 
Differentiating Eq. (\ref{omegadevelop}) 
with respect to $x^{\gamma}$, and employing 
Eqs. (\ref{Riccilemma}), (\ref{Gammaderiv1})
and (\ref{Gammaderiv2}) implies that 
\begin{eqnarray}
\label{bigequation}
(\nabla_{\kappa} \Omega^{\kappa}_{\mu})^* = 
\Omega^{\eta}_{\kappa}(\phi\Omega^{\kappa}_{\eta;\mu} 
+ \Omega^{\kappa}_{\eta} \nabla_{\mu} \phi ) - 
\Omega^{\kappa}_{\mu}(\Omega \nabla_{\kappa} \phi + 
\phi\nabla_{\kappa} \Omega ) + \nabla_{\mu} (\nabla^2 \phi) 
\nonumber \\ 
+ \Omega\Omega^{\kappa}_{\mu} \nabla_{\kappa} \phi -
\frac{2\Lambda\epsilon}{n - 1}\nabla_{\mu} \phi  + 
\phi ( \Omega^{\kappa}_{\mu} \nabla_{\kappa} \Omega + 
\Omega \nabla_{\kappa} \Omega^{\kappa}_{\mu} - \epsilon 
\nabla_{\lambda} R^{\lambda}_{\mu}) .
\end{eqnarray}

The trace of Eq. (\ref{omegadevelop}), on the other hand,  
is given by 
\begin{equation}
\label{Omegastar}
\Omega^* = \epsilon \nabla^2 \phi 
-\epsilon \phi \left( R+\frac{2n\Lambda}{n-1} \right) +\phi \Omega^2
\end{equation}
and combining Eq. (\ref{bigequation}) 
with the covariant derivative of Eq. (\ref{Omegastar}) 
with respect to $x^{\mu}$ then implies that 
\begin{eqnarray}
\label{largeequation}
\left( \nabla_{\lambda} \Omega^{\lambda}_{\mu} 
- \nabla_{\mu}\Omega \right)^* = 
\nabla_{\mu} \phi \left ( \epsilon(R + 2\Lambda) + 
\Omega^{\eta}_{\kappa}\Omega^{\kappa}_{\eta} - 
\Omega^2 \right) \nonumber \\
+ \phi \left (\Omega^{\eta}_{\kappa} \nabla_{\mu} \Omega^{\kappa}_{\eta} + 
\Omega \nabla_{\lambda} \Omega^{\lambda}_{\mu} - 2\Omega
\nabla_{\mu} \Omega + 
\epsilon \nabla_{\mu} R - \epsilon \nabla_{\lambda}
R^{\lambda}_{\mu} \right )  .
\end{eqnarray}
The first bracketed term on the right hand side 
of Eq. (\ref{largeequation})
vanishes 
due to Eq. (\ref{gauss}). The second bracketed term vanishes 
due to the covariant derivative of Eq. (\ref{gauss}) 
with respect to $x^{\mu}$ and 
the contracted Bianchi identity 
\begin{equation}
\nabla_{\lambda}
R^{\lambda}_{\mu} = \frac{1}{2} \nabla_{\mu}R   .
\end{equation}
Hence, Eq. (\ref{deriv1}) 
is valid and Eq. (\ref{codazzi}) propagates. 

In establishing the validity of Eq. (\ref{deriv2}), 
it is necessary to calculate $R^*$. Since $R$ is a scalar, 
its derivative  
can be evaluated in normal coordinates \cite{campbell}. 
Employing 
Eqs. 
(\ref{codazzi}),  (\ref{rielem}) 
and (\ref{Gammaderiv2})
implies that 
\begin{equation}
\label{Rstar}
R^* = 2\Omega \nabla^2 \phi +2 R_{\alpha\beta} \Omega^{\alpha\beta}
\phi -2\Omega^{\gamma}_{\beta} 
\nabla_{\gamma}^{\beta} \phi .
\end{equation}
However, it follows from Eq. (\ref{omegadevelop}) and (\ref{Omegastar}) 
that
\begin{eqnarray}
\label{gau2}
(\Omega^{\mu}_{\lambda}\Omega^{\lambda}_{\mu} - \Omega^2)^* =
2\Omega\phi \left( \Omega^{\mu}_{\lambda}\Omega^{\lambda}_{\mu} 
- \Omega^2 + \epsilon(R + 2\Lambda) \right) \nonumber \\
- 
2\epsilon \left( \phi\Omega^{\mu}_{\lambda}R^{\lambda}_{\mu} 
+ \Omega \nabla^2 \phi - \Omega^{\mu}_{\lambda} 
\nabla^{\lambda}_{\mu}  \phi
\right)   .
\end{eqnarray}
Thus, substitution of Eqs. (\ref{gauss}) and 
(\ref{Rstar}) into Eq. (\ref{gau2}) implies that 
Eq. (\ref{deriv2}) is valid.

\end{document}